\documentclass[nofootinbib,amsmath,amssymb,floatfix,]{WileyMSP-template}

\usepackage{graphicx}
\usepackage{dcolumn}
\usepackage{bm}
\usepackage{physics}
\usepackage{xcolor}
\usepackage{booktabs} 
\usepackage{caption}
\usepackage{subcaption}
\usepackage[hidelinks]{hyperref}
\usepackage[utf8]{inputenc}
\usepackage[T1]{fontenc}
\usepackage[square,numbers]{natbib}
\begin{document}


\title{Dynamic control of photon-magnon interactions via secondary magnon excitation}

\maketitle

\author{Fizaan Khan\footnote{fizaan.khan.phy21@iitbhu.ac.in}}
\author{Sachin Verma}
\author{Biswanath Bhoi}
\author{Rajeev Singh\footnote{rajeevs.phy@iitbhu.ac.in}}



\begin{affiliations}
Nano-Magnetism and Quantum Technologies Laboratory\\
Indian Institute of Technology (BHU), Varanasi\\
221005 - India
\end{affiliations}



\begin{abstract}

Photon-mediated magnon-magnon coupling between spatially separated Yttrium Iron Garnet (YIG) and permalloy (NiFe) thin films on a planar hexagonal ring resonator shows clear signatures of magnon-magnon interaction are observed without direct dipolar interaction between the magnetic films. 
The coupling strength between the hexagonal ring resonator and the permalloy film increases with the thickness of the YIG film, despite a fixed permalloy film thickness. 
This suggests the presence of an indirect interaction channel mediated by resonator photons. 
A theoretical model is presented that accurately reproduces the observed transmission spectra and reveals a nontrivial interdependence between the individual coupling strengths of YIG and permalloy to the resonator. 
These results highlight the importance of indirect interactions and potential crosstalk pathways in designing hybrid magnonic systems and scalable quantum architectures, while demonstrating the feasibility of cost-effective, planar configurations for experimental implementation. 
These insights are valuable for advancing low-loss, coherent information transfer in hybrid quantum devices.

\end{abstract}


\begin{figure*}
    \centering
    \begin{subfigure}[b]{.45\textwidth}
        \includegraphics[width=\linewidth]{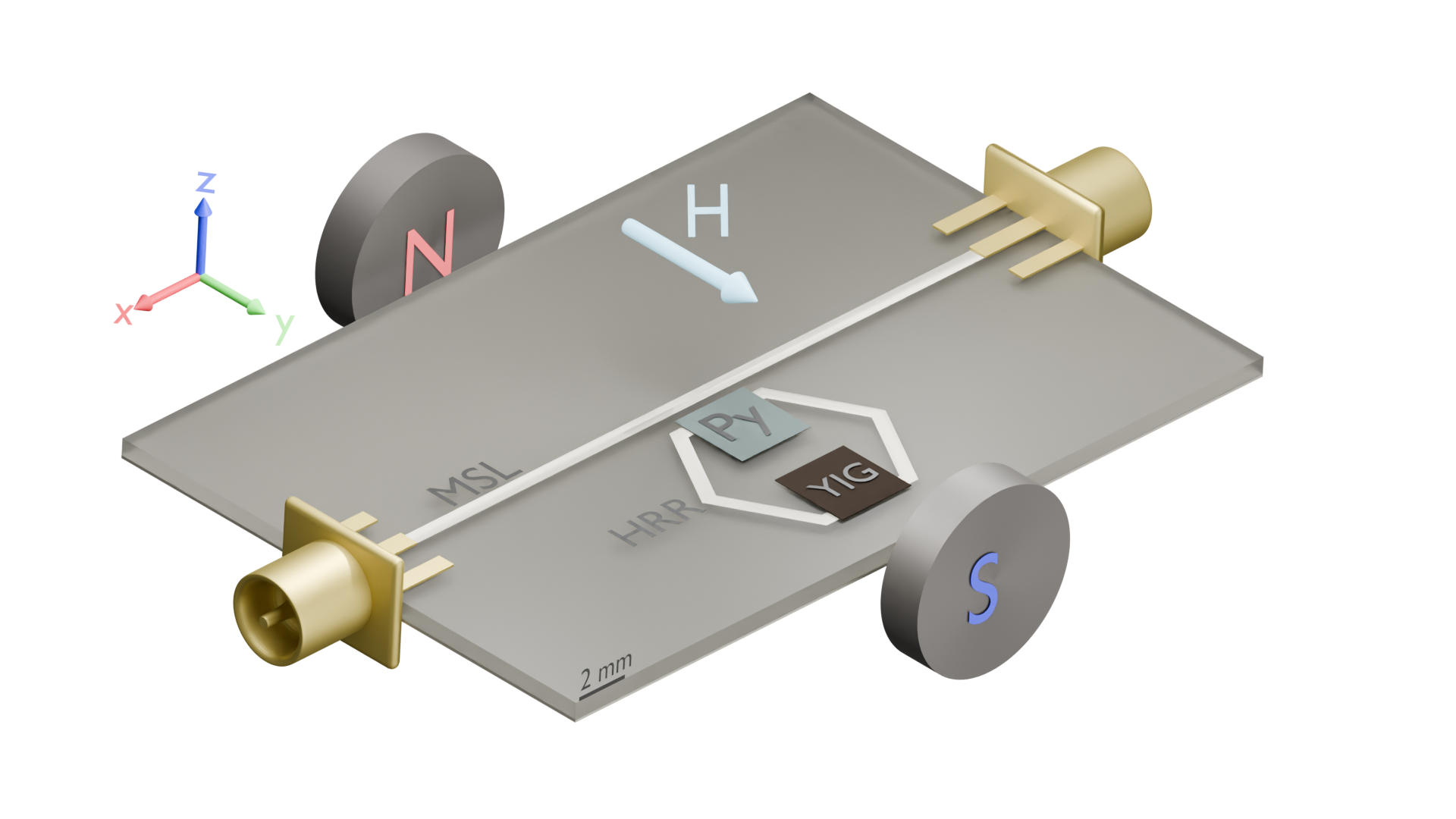}
    \caption{\label{fig:3d-schematic}}
    \end{subfigure}
    \begin{subfigure}[b]{.45\textwidth}
        \includegraphics[width=\linewidth]{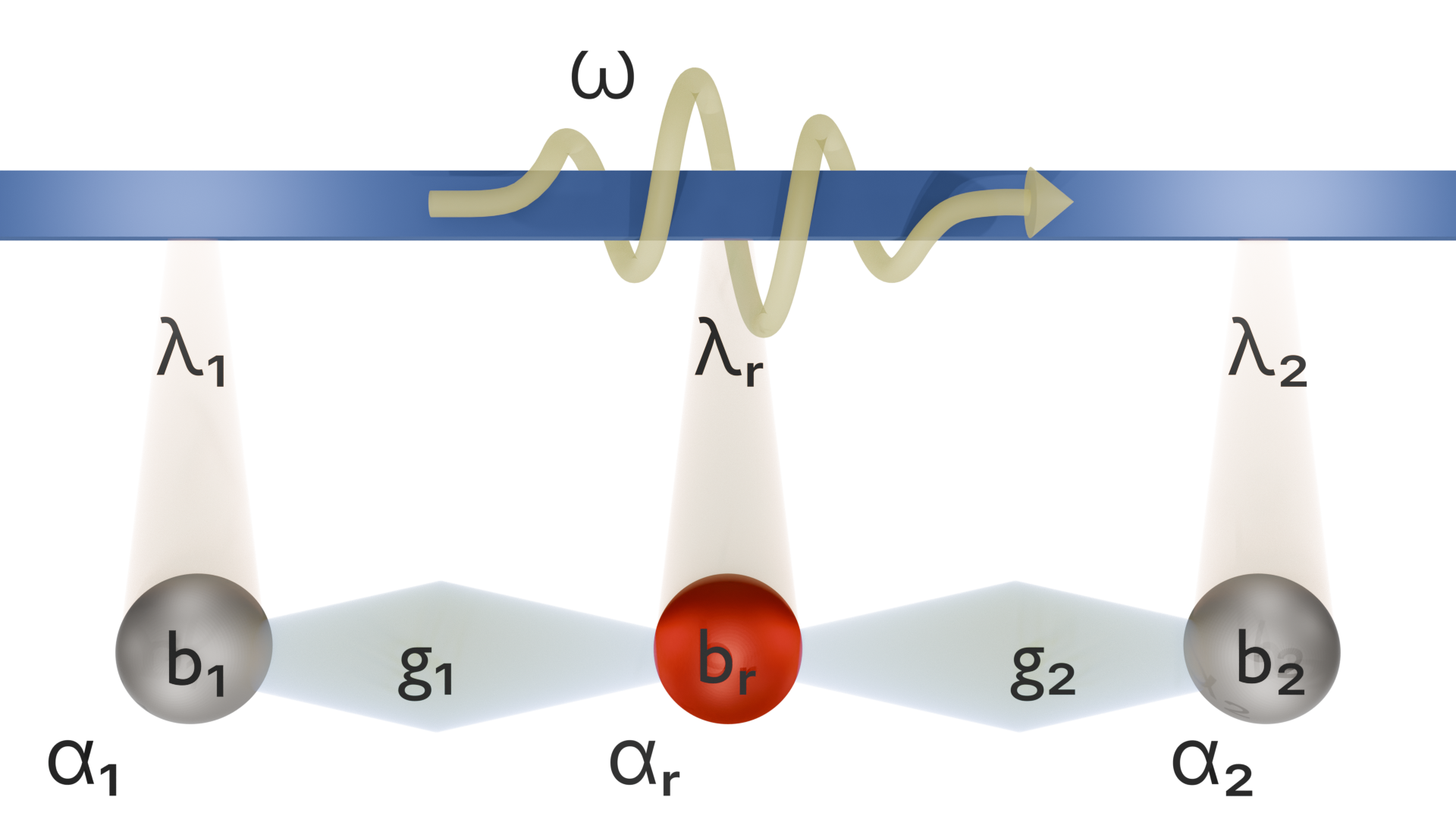}
    \caption{\label{fig:model-cartoon}}
    
    \end{subfigure}
    \caption{(a) The setup includes the two magnonic films placed on a pair of opposing sides of a copper hexagon (the photonic resonator). The ports are plugged into a vector network analyzer to both excite the microstrip line and also measure the transmission spectrum while a magnetic field is applied along the \(y\)-direction. (b) is the schematic of the model developed for this system. The spheres are bosonic oscillators with creation (annihilation) operator \(\hat{b}_j^\dagger\)(\(\hat{b}_j\)). \(\alpha_j\) denote intrinsic damping of the oscillators, and \(\lambda_j\) couple the oscillators to the traveling photon. The constants \(g_1\) and \(g_2\) denote the coupling of the magnonic cavities to the photonic cavity.}
    \label{fig:3d-and-model}
\end{figure*}

\section{Introduction}
The advancement of hybrid quantum systems that integrate magnonic and photonic components has emerged as a critical frontier in the pursuit of scalable quantum information processing and next-generation microwave technologies\cite{Wagle_2024,zhao2022,zhu2020,zhang2021,fan2025}. 
In such systems, magnons—the quanta of spin waves—interact coherently with microwave photons, enabling unique functionalities such as tunable coupling\cite{li2023tunablemagnonphotoncouplingmagnon,Ghirri2023}, non-reciprocity\cite{Kim2024,freeman2025,cosset2024,chakraborty2023}, and quantum transduction\cite{nongthombam2023}. 
As these device architectures continue to scale toward nanoscale dimensions \cite{hou2019,li2019}, the complexity of electromagnetic interactions increases, particularly due to the emergence of indirect or unintended couplings between spatially separated components. 
Shared photonic modes or circuit environments often mediate these couplings and can lead to both detrimental crosstalk and useful long-range interactions \cite{Xu2024,Jeon2024}. 
Understanding and controlling these mediated interactions is essential for designing scalable hybrid quantum devices that are also coherent and robust.

Material choice plays a crucial role in determining coherence, coupling strength, and scalability in hybrid quantum systems. Yttrium Iron Garnet (YIG), with its low damping and long spin-wave coherence, has emerged as a leading candidate for magnonic quantum circuits \cite{Wagle_2024,Ghirri2023,Bhoi2017,Andrich_2017}. 
It enables strong and ultrastrong photon-magnon coupling in planar superconducting resonators, with recent studies\cite{Ghirri2023} showing coupling-to-frequency ratios \(>\) 0.2. 
Likewise, lithographically patterned organic ferrimagnets have demonstrated cooperativity values exceeding \(10^3\) in the hybrid quantum modes \cite{Xu2024}, paving the way for their integration with superconducting qubit platforms and enabling circuit quantum electrodynamics (cQED) functionalities within magnon-based architectures. 
These advances highlight the growing maturity of hybrid photon-magnon coupled (PMC) systems. 
They underscore the need for deeper exploration into their nontrivial coupling dynamics, particularly those arising from indirect, photon-mediated interactions between spatially separated magnetic elements.

Unresolved theoretical and experimental challenges persist, particularly in understanding mediated coupling, the dependence of interaction strength on magnetization and spatial configuration, and the role of decoherence in nonlocal systems. 
Notably, Hyde \textit{et al.} \cite{cannot-g3} observed that specific indirectly coupled modes can surpass their directly coupled counterparts in transmission amplitude, highlighting the need to reassess the conventional assumptions surrounding circuit isolation and crosstalk. 
As previous work suggests that cavity photons can act as effective mediators for long-distance coupling in hybrid architectures \cite{grigoryan2019}, we investigate this phenomenon using a hexagonal ring resonator as the coupling platform. 
It is planar and symmetric: well-suited for probing such mediated interactions and exploring their impact on the coherent dynamics of spatially separated magnetic elements.

This study demonstrates that photon-mediated interactions between a YIG film and a spatially separated permalloy film, integrated via the resonator, can give rise to measurable magnon-magnon coupling, even in the absence of significant direct dipolar overlap. 
Our choice of configuration is in contrast with systems often studied, such as YIG spheres\cite{morris2017,zhang2016,bourhill2016,Dilawaiz_2024} and YIG/permalloy bilayers\cite{inman2022,Santos2023,Xiong2020} that carry direct coupling.
The observed coupling arises from the nonlocal electromagnetic modes of the resonator, which effectively bridge the two magnetic films. 
Importantly, the coupling strength is found to depend sensitively on the geometric configuration and electromagnetic response of circuit elements that are otherwise considered independent\cite{cannot-g3}.
Notably, this mediated interaction persists even when the individual photon-magnon coupling lies well below the ultrastrong coupling threshold, emphasizing the relevance of indirect pathways in the overall system dynamics. 
These results suggest that conventional approaches to suppressing crosstalk in magnonic integrated circuits may need to be revised to account for long-range electromagnetic interactions. 
To model the system, we employ input–output theory\cite{walls-2007,shrivastava-2024} instead of the conventional Schrieffer–Wolff transformation\cite{Grigoryan_2013,schrieffer1966}, allowing us to extract (\(S_{21}\)) and directly compare theoretical predictions with simulation data. 
The design is intentionally simple and scalable: implemented on a millimeter scale, operable at room temperature. 
Despite its simplicity, the platform offers a promising testbed for investigating mediated coupling in accessible, low-cost quantum device architectures.

\section{Design and Numerical Modeling of the Hybrid Photon-Magnon System}

\begin{figure}
    \centering
    \includegraphics[width=\linewidth]{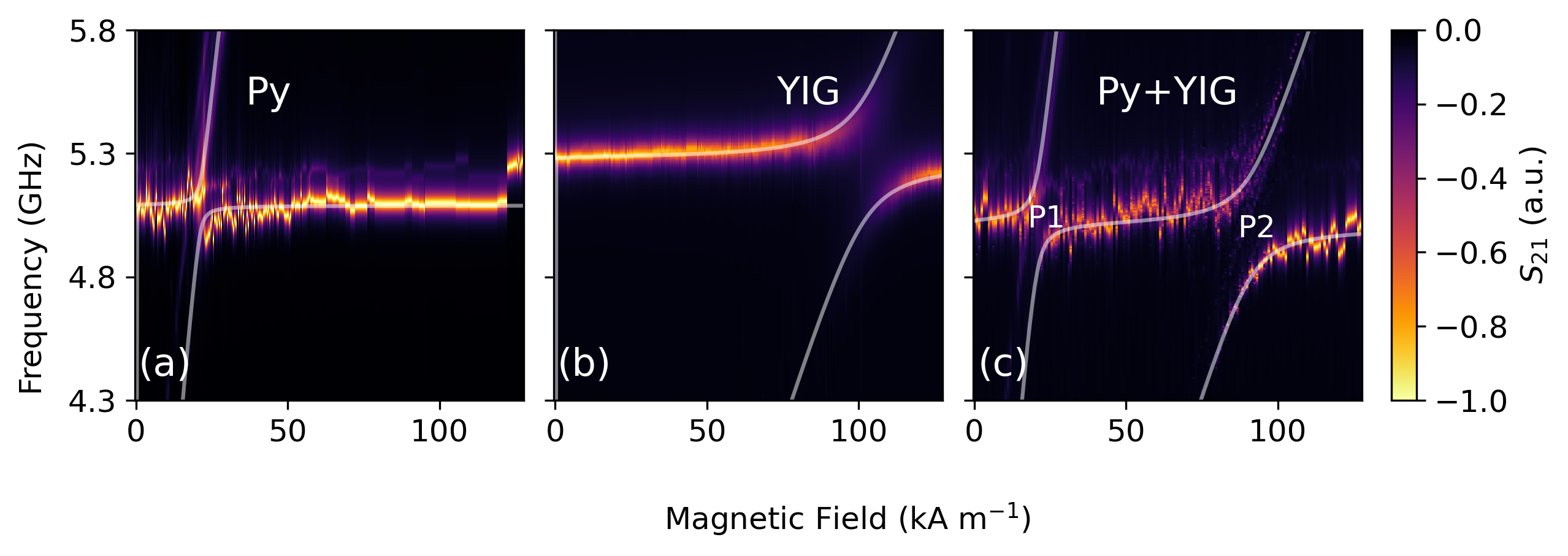}
    \caption{A comparison of the magnonic coupling strengths of the (a) permalloy, (b) YIG, and (c) both to the resonator in an indirectly coupled configuration. The overlaid white curves are the real part of the eigenvalues of (\ref{eq:coupling-matrix}) at each value of \(H\) and was used to fit the model.}
    \label{fig:py-yig-comb}
\end{figure}
In designing hybrid systems, it is essential to account for the various forms of hybridization and their collective influence on the system’s dynamics. 
We investigate magnonic interactions in a configuration specifically chosen to minimize direct dipolar coupling. 
This allows us to isolate and analyze interaction pathways mediated indirectly through a standard bosonic mode, i.e., photons.
The observed hybrid system features the resonator parked near a microstrip line that is 0.57 mm wide and 30 mm long. It integrates with the two 3 mm \(\times\) 3 mm magnetic films—YIG and permalloy—mounted on opposite arms of the resonator within a planar geometry, as illustrated in Figure \ref{fig:3d-schematic}. 
When a microwave current propagates along the \(x\)-axis (see axes in Figure \ref{fig:3d-schematic}) through the microstrip line (labelled `MSL' in Figure \ref{fig:3d-schematic}), it produces a transverse microwave magnetic field that excites the hexagonal resonator (labelled `HRR'). 
It functions as a parallel-LC circuit exhibiting quasi-static resonance. 
This resonant field, in turn, interacts with the magnetic YIG (black film in Figure \ref{fig:3d-schematic}) and permalloy (of thickness 20 \(\mu\)m, dirty green film in the Figure) films placed on the resonator, inducing their respective magnon modes and enabling the probing of respective photon–magnon coupling dynamics. 
The high-frequency electromagnetic solver CST Studio Suite was employed to simulate individual hybrid configurations: resonator–permalloy, resonator–YIG, and the combined hybrid system resonator–YIG–permalloy. 

To investigate photon-mediated magnon–magnon coupling, we obtained \(S_{21}\) as a function of microwave frequency (\(\omega\)) under a static external magnetic field (\(H\)) applied along the \(y\)-axis at room temperature. 
Figures \ref{fig:py-yig-comb}a, \ref{fig:py-yig-comb}b, and \ref{fig:py-yig-comb}c show the \(S_{21}\) power spectra on the \(\omega\)–\(H\) plane for the hybrid resonator–permalloy, resonator–YIG, and resonator–YIG–permalloy, respectively. 
In this setup, the photon mode associated with the resonator behaves as a fixed-frequency resonance determined by the geometry of the LC structure. 
In contrast, the magnon modes in YIG and permalloy follow the Kittel relation\cite{kittel1948}, resulting in a resonance that shifts with the applied magnetic field. 
The strength of photon-magnon coupling is inferred from the degree of avoided level crossing--an established indicator of coherent energy exchange--and is a particularly relevant parameter in applications such as quantum sensing and quantum memory buses\cite{Xu2024}.

As shown in Figure \ref{fig:py-yig-comb}a, the permalloy–resonator system exhibits only weak coupling, which is expected due to the metallic nature of permalloy that introduces higher damping and noise. 
In our model, we observe a coupling constant of \(g=0.11\) corresponding to this interaction.
In contrast, Figure 2b demonstrates that the YIG–resonator system exhibits strong coupling (\(g=0.25\)), attributable to the low damping and high spin-wave coherence of the magnetic material involved.
Outside the influence of the magnonic film's resonance behavior, the photonic resonator returns to its natural frequency, observed in the rise of the horizontal line in Figure \ref{fig:py-yig-comb}b.

Interestingly, in the combined system (Figure \ref{fig:py-yig-comb}c), the permalloy mode displays a markedly enhanced coupling signature compared to its isolated behavior, while the YIG coupling appears slightly diminished.
We note that fitted values are now \(g_1=0.2\) (of permalloy) and \(g_2=0.21\) (of YIG), a simultaneous enhancement of permalloy-resonator coupling (seen at crossing marked P1 in Figure \ref{fig:py-yig-comb}c) and reduction of YIG-resonator coupling (crossing P2).
This suggests a redistribution of interaction strength facilitated by the resonator’s photon mode, emphasizing the role of photon mediation in enabling effective magnon–magnon interactions between spatially separated magnetic elements.

\section{Theoretical Formalism for Photon Mediated Magnon-Magnon Coupling}\label{section:theory}

We developed a microscopic model based on quantum mechanical principles to gain deeper insight into the numerically observed photon-mediated magnon–magnon coupling and quantitatively estimate the corresponding interaction strength. 
This approach provides a robust framework for describing the coupled dynamics of the system, effectively capturing both direct and indirect interactions. 
It is readily extended to systems involving multiple coupled quantum harmonic oscillators.

Figure \ref{fig:model-cartoon} attempts to visually illuminate the nature of this model.
Three oscillators named 1, 2, and \(r\), each with its own internal damping and a coupling to travelling photons, interact with one another.
While oscillators 1 and 2 interact with oscillator \(r\) directly, they are not allowed to interact with one another.

The system Hamiltonian for such a model is best understood as a sum of three parts, 
\(
{\hat{H}}={\hat{H}}_{\text{o}}+{\hat{H}}_{\text{msl}}+{\hat{H}}_{\text{int}}\).
The natural Hamiltonian \(\hat{H}_{\text{o}}\), is the familiar sum of quantum harmonic oscillators,
\begin{equation}
    {\hat{H}}_{\text{o}}/\hbar = \sum_{j \in \{r,1,2\}} \tilde{\omega}_j \hat{b}_j^\dagger \hat{b}_j
    \label{eq:non-int-ham}
\end{equation}
where \(\tilde{\omega}_j=\omega_j-i\alpha_j\) refers to the damped natural frequency of the oscillator for a resonant frequency \(\omega_j\) and an intrinsic damping constant \(\alpha_j\). 
For the index \(j\), we use \(r\) to denote the photonic resonator and \(1\) and \(2\) to denote the permalloy and YIG magnons, respectively.

We then provide a driving field through a micro stripline, which is assigned the annihilation (creation) operator \(\hat{p}_k\) (\(\hat{p}_k^\dagger\)) for the \(k\)th mode:
\begin{equation}
    \hat{H}_\text{msl}/\hbar = \int \omega_k \hat{p}_k^\dagger \hat{p}_k \, dk + \sum_{j \in \{r,1,2\}} \int  \lambda_j \left( \hat{b}_j^\dagger + \hat{b}_j \right) \left( \hat{p}_k^\dagger + \hat{p}_k \right) \, dk
    \label{eq:msl-ham}
\end{equation}
where \(\hat{b}_j\) (\(\hat{b}_j^\dagger\)) are the bosonic annihilation (creation) operators.
We introduce \(\lambda_j\) to denote the coupling of the oscillators to the stripline.

The final term in the Hamiltonian is the mutual interaction among all the resonators, photonic and magnonic.
The coupling constants \(g_j\) that populate \(\hat{H}_{\text{int}}\) are of primary interest to us. 
This part of the Hamiltonian is given  by
\begin{equation}
    \hat{H}_\text{int}/\hbar= g_1 \hat{b}_1^\dagger \hat{b}_r + g_2 \hat{b}_2^\dagger \hat{b}_r+\text{h.c.}
    \label{eq:int-ham}
\end{equation}
and captures the crosstalk between the resonator modes, leading to the observed normal anticrossing in Figure \ref{fig:py-yig-comb}.
The normal anticrossing behaviour is due to real coupling constants in the Hamiltonian \cite{bernier2018,metelmann2014}.

The equations of motion for the three oscillators are readily calculated by the standard Heisenberg-Langevin formalism \cite{rao2020,shrivastava-2024}.
Following a method similar to that followed by Srivastava \textit{et al.}\cite{shrivastava2024emergencecouplinginducedtransparency}we set \(\beta_i=2\pi\lambda_i^2\), the \(S_{21}\) parameter for such a system, given by \(S_{21}=\hat{p}_{\text{out}}/\hat{p}_{\text{in}}-1\) is
\begin{equation}
    S_{21} = \frac{2}{i} \sum_{j}\sqrt{\beta_j} \frac{\hat{b}_j (\omega)}{\hat{p}_{\text{in}} (\omega)}
    \label{eq:s21-analytical}
\end{equation}
which, rewritten in the form of a matrix
\begin{equation}
S_{21} = B^T \mathcal{M}^{-1} B
\label{eq:s21_matrix}
\end{equation}
where
\[
B = \sqrt{2} 
\begin{bmatrix}
\sqrt{\beta_1} \\
\sqrt{\beta_r} \\
\sqrt{\beta_2}
\end{bmatrix},
\]
and \(\mathcal{M} = i\left(\omega\mathcal{I}-\hat{H}_{\text{coupling}}\right)\), where \(\mathcal{I}\) is the \(3\times3\) identity matrix and \(\hat{H}_{\text{coupling}}\) is the effective coupling Hamiltonian, that solves the equations of motion for the mode operators, given by
\begin{equation}
    \hat{H}_{\text{coupling}} = 
\begin{bmatrix}
 \tilde{\omega}'_1 & g_{1} - i\sqrt{\beta_1 \beta_r} & -i\sqrt{\beta_1 \beta_2} \\
g_{1} - i\sqrt{\beta_1 \beta_r} &  \tilde{\omega}'_r & g_{2} - i\sqrt{\beta_r \beta_2} \\ -i\sqrt{\beta_1 \beta_2} & g_{2} - i\sqrt{\beta_r \beta_2} &  \tilde{\omega}'_2
\end{bmatrix}.
\label{eq:coupling-matrix}
\end{equation}

The terms on the diagonal include intrinsic and extrinsic damping terms, in the form \(\tilde{\omega}'_j= \tilde{\omega}_j-i\beta_j\)\cite{shrivastava2024emergencecouplinginducedtransparency}. 
Since we are ignoring terms of direct magnon coupling, there are no coupling constants on the edges of the off-diagonal.
The parameters \(g_1\) and \(g_2\) can be obtained by fitting the model to observations at each \(H\). 

\section{Discussion}

\begin{figure*}[ht!]
    \centering
    \includegraphics[width=\textwidth]{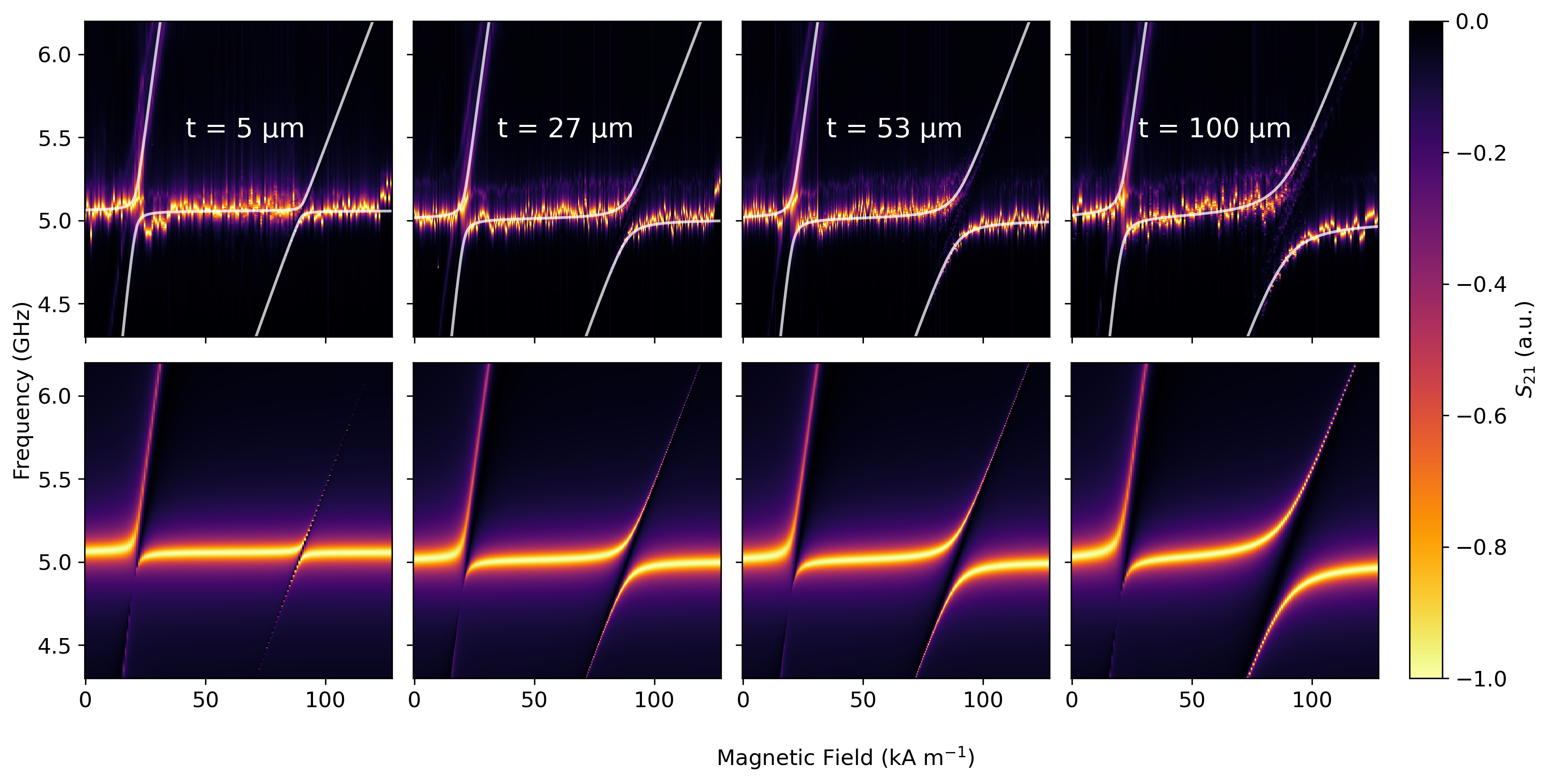}
    \caption{Observed (top row) and calculated (bottom row) \(S_{21}\) spectra of the system for corresponding selected thicknesses of YIG. The permalloy crossing is also seen to widen with a change in YIG thickness.} 
    \label{fig:cst-double-film-t-sweep}
\end{figure*}

\begin{figure*}
    \centering
    \includegraphics[width=\linewidth]{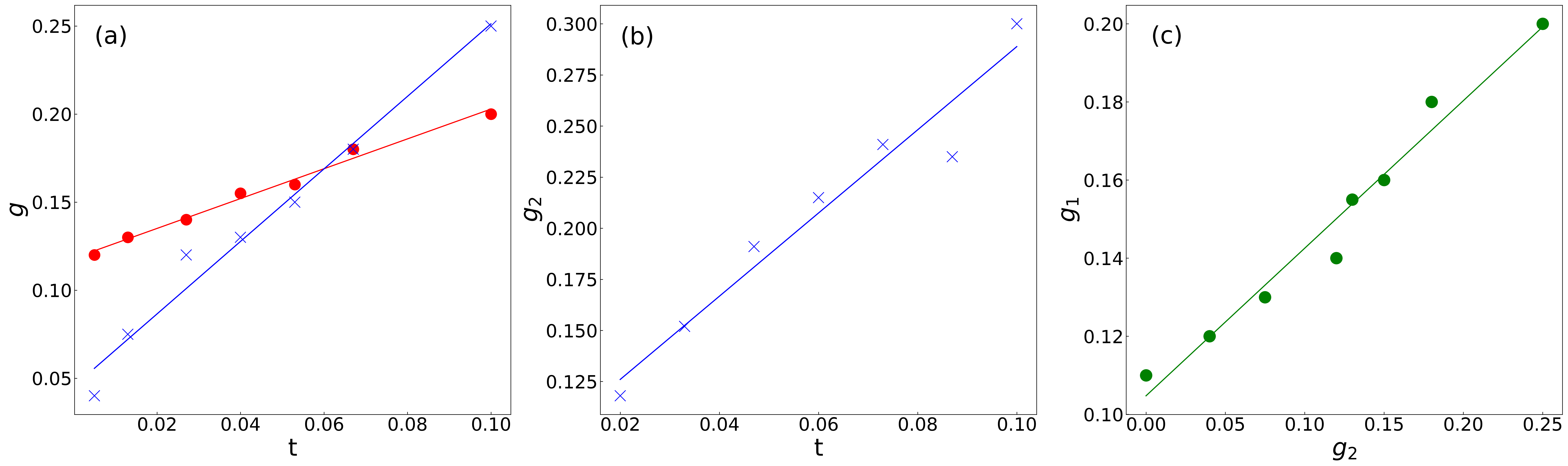}
    
    \caption{The blue crosses in (a) and (b) represent the coupling strengths of the YIG film to the resonator. The coupling strengths in (a) are for a system that does not include a permalloy film. Upon introduction of the permalloy in (b), the coupling strengths of the YIG have decreased by a fixed amount. The red dots in (b) mark \(g_1\) (of permalloy), the coupling strengths of the permalloy to the resonator, and blue crosses are \(g_2\) (of YIG), the coupling strengths of the YIG to the resonator. In (c), we plot the linear relationship between the two coupling strengths, \(g_1(g_2)\).}
    \label{fig:couplings-grand}
\end{figure*}

Choosing to exclude terms of direct coupling in the interaction Hamiltonian (\ref{eq:int-ham}) was far from an arbitrary one.
The configuration of the system is such that we observe minimal overlap between the two magnonic films on account of the close-to-planar thin films.
This behavior is mirrored by the model, which accurately predicts the \(S_{21}\) spectrum in the absence of a direct coupling term.

The coupling constants predictably affect the extent of avoidedness of the avoided crossing for the corresponding magnon (identified by the coupling centers), with greater constants leading to greater avoidedness.

To investigate the presence of indirect magnon–magnon coupling mediated by cavity photons, we performed a series of simulations by varying the thickness (\(t\)) of the YIG film (from \(t = 5\) to \(100\) \(\mu\)m) while keeping the properties of the permalloy film and the hexagonal microwave resonator unchanged. 
Increasing the YIG thickness enhances its magnetization, thereby modifying its magnonic response.

As shown in the top row of Figure \ref{fig:cst-double-film-t-sweep}, an increase in the mode splitting between the upper and lower branches (for both regions P1 and P2) of the magnon–photon hybrid modes is observed with thicker YIG films. 
Notably, this occurs when the permalloy film remains unaltered, indicating that the enhanced coupling is not due to changes in itself.
We make a more direct observation of this in the plots in Figure \ref{fig:couplings-grand}, where we compare the coupling of the resonator-YIG in the absence of the permalloy film in Figure \ref{fig:couplings-grand}a and in its presence in Figure \ref{fig:couplings-grand}b (blue crosses).
Both linear.

Evidently, while the slope of the resonator-YIG coupling (i.e., \(g_2\) vs \(t\)) remains unchanged from Figure \ref{fig:couplings-grand}b to Figure \ref{fig:couplings-grand}a, the intercept does drop slightly.
The presence, then, of the permalloy film causes a decrease in resonator-YIG coupling independent of the thickness of the YIG film.
Combined with the non-zero slope of the unvarying resonator-permalloy couple, the behavior provides strong evidence for photon-mediated interaction between the YIG and permalloy magnon modes, a hallmark of indirect magnon–magnon coupling through the shared cavity photon field.

To validate these observations, we used a theoretical model to compute the transmission spectrum (\(S_{21}\)) under identical conditions. The bottom row of Figure \ref{fig:cst-double-film-t-sweep} presents the calculated spectra, which closely match the corresponding observed results in the top row. 
The evolution of the spectra with increasing YIG thickness—from left to right—demonstrates consistent hybrid dynamics, affirming the model’s accuracy. 
Predictably, as YIG becomes thicker, its coupling to the resonator increases due to the higher magnetization. 
However, a notable and somewhat counterintuitive result is the simultaneous increase in the coupling strength of the permalloy mode, despite its parameters remaining fixed. 
This reveals that the YIG film exerts an indirect influence on the permalloy–resonator coupling, mediated by their mutual interaction with the cavity mode.
We can ascribe to the resonator-permalloy coupling a dependence on the resonator-YIG coupling, \(g_1(g_2)\), which we observe in this work to be linear (Figure \ref{fig:couplings-grand}c).



\section{Conclusions}
In this study, we have demonstrated and quantified the photon-mediated coupling between two spatially separated magnonic films, YIG and permalloy, integrated onto a planar hexagonal ring resonator. 
Our experimental results reveal that even without direct dipolar interaction, a clear magnon-magnon coupling emerges, evidenced by the anticrossing in the transmission spectra. 
A key finding is the non-trivial enhancement of the permalloy-resonator coupling strength as the thickness of the YIG film increases, while the permalloy film itself remains unchanged. 
This observation strongly indicates that the resonator's photonic mode acts as a quantum bus, facilitating an indirect interaction pathway between the two magnonic systems.

Furthermore, the theoretical model developed using the input-output formalism accurately reproduces the experimental transmission spectra across various YIG thicknesses. 
The model confirms a linear relationship between the coupling strengths of the YIG and permalloy films to the resonator, quantitatively capturing the indirect influence one magnon has on the other. 
These findings underscore the critical role of mediated interactions in hybrid magnonic systems, offering a new mechanism for controlling coupling strengths in scalable, on-chip architectures and paving the way for advanced quantum and microwave technologies.


\medskip

\textbf{Acknowledgements} \par 
The work was supported by the Council of Science and Technology, Uttar Pradesh (CSTUP) (Project Id: 2470, CST, U.P. sanction No: CST/D-1520 and Project Id: 4482, CST, U.P. sanction No: CST/D-7/8).
B. Bhoi acknowledges support from the Science and Engineering Research Board (SERB) India - SRG/2023/001355. 
S. Verma acknowledges the Ministry of Education, Government of India, for the Prime Minister’s Research Fellowship (PMRF ID-1102628).

\section*{Statements and Declarations}

\textbf{Conflict of Interest:} The authors hold no conflicts of interest. 

\textbf{Author Contributions:} All authors contributed to the study’s conception and design. R.S. and B.B. led the work and wrote the manuscript with F. Khan. The other co-authors read, commented, and approved the final manuscript.  

\textbf{Data Availability:}
The data supporting this study’s findings are available within the article.

\section*{Appendix}
The following is a detailed derivation of the transmission parameter \(S_{21}\), using the input-output formalism\cite{walls-2007,shrivastava2024emergencecouplinginducedtransparency}.

The Hamiltonian of the system was asserted in Section \ref{section:theory} to exist in three parts: (\ref{eq:non-int-ham}), (\ref{eq:msl-ham}), and (\ref{eq:int-ham}). Plugging these into \(\hat{H}=\hat{H}_\text{o}+\hat{H}_\text{msl}+\hat{H}_\text{int}\) and making the rotating wave approximation we obtain
\begin{equation}
\hat{H}/\hbar=\sum_{j \in \{r,1,2\}}\left[\tilde{\omega}_{j}\hat{b}_{j}^{\dagger}\hat{b}_{j}+\int_{-\infty}^{\infty}\lambda_{j}(\hat{b}_{j}^{\dagger}\hat{p}_{k}+\hat{p}_{k}^{\dagger}\hat{b}_{j})dk\right]+[g_{1}\hat{b}_{1}^{\dagger}\hat{b}_{r}+g_{2}\hat{b}_{2}^{\dagger}\hat{b}_{r}+h.c.] +\int_{-\infty}^{\infty} \omega_{k}\hat{p}_{k}^{\dagger}\hat{p}_{k}dk
\end{equation}

The dynamics of the system can be described by the Heisenberg-Langevin equations of motion, which are derived from the Hamiltonian and the fundamental commutation relations for bosonic operators that have the commutation relations \([\hat{p}_{k'}^{\dagger}, \hat{p}_{k}] = \delta(k' - k)\) and \([\hat{b}_{i}^{\dagger}, \hat{b}_{j}] = \delta_{i,j}\).

The solutions to the Heisenberg equations of motion for \(\hat{p}_k(t)\) and \(\hat{p}_k^\dagger(t)\) for some \(t_0<t\) are the following:
\begin{subequations}
\begin{align}
\hat{p}_k(t) &= e^{-i\omega_k(t-t_0)}\hat{p}_k(t_0) - i\sum_j\lambda_j  \int_{t_0}^{t} \hat{b}_j(t') e^{-i\omega_k(t-t')} dt' \\
\hat{p}^\dagger_k(t) &= e^{-i\omega_k(t-t_0)}\hat{p}^\dagger_k(t_0) - i\sum_j \lambda_j \int_{t_0}^{t} \hat{b}^\dagger_j(t') e^{-i\omega_k(t-t')} dt'
\end{align}
\label{eq:input-cont}
\end{subequations}
Whereas, when the solution is sought until some \(t_1>t\):
\begin{subequations}
\begin{align}
\hat{p}_k(t) &= e^{-i\omega_k(t-t_1)}\hat{p}_k(t_1) +i\sum_j\lambda_j  \int_{t}^{t_1} \hat{b}_j(t') e^{-i\omega_k(t-t')} dt' \\
\hat{p}^\dagger_k(t) &= e^{-i\omega_k(t-t_1)}\hat{p}^\dagger_k(t_1) + i\sum_j \lambda_j \int_{t}^{t_1} \hat{b}^\dagger_j(t') e^{-i\omega_k(t-t')} dt'
\end{align}
\label{eq:output-cont}
\end{subequations}
We pull the \(\lambda_j\) out of the integral under the assumption that the transfer of energy between the operators \(\hat{p}_k\) and \(\hat{b}_j\) is a Markov process. 
For \(\hat{b}_j\):
\begin{subequations}
\begin{align}
\dot{\hat{b}}_{1,2}(t) &= -i\tilde{\omega}_{1,2}\hat{b}_{1,2}(t) - i\lambda_{1,2}\int \hat{p}_k(t) dk - ig_{1,2}\hat{b}_r(t) \\
\dot{\hat{b}}_r(t) &= -i\tilde{\omega}_r \hat{b}_r(t) - i\lambda_r \int \hat{p}_k(t) dk - i(g_1^* \hat{b}_1 + g_2^* \hat{b}_2)(t).
\end{align}
\label{eq:resonator-modes}
\end{subequations}

We will then define \(\hat{p}_{\text{in}}(t)\) and \(\hat{p}_{\text{out}}(t)\) as
\begin{subequations}
\begin{align}
\hat{p}_{\text{in}}(t) &= \frac{1}{\sqrt{2\pi}} \int e^{-i\omega_k(t-t_0)} \hat{p}_k(t_0) dk \\
\hat{p}_{\text{out}}(t) &= \frac{1}{\sqrt{2\pi}} \int e^{-i\omega_k(t-t_1)} \hat{p}_k(t_1) dk.
\end{align}
\label{eq:input-output-modes}
\end{subequations}

We can now define the decay rate \(\beta_j=2\pi\lambda_j^2\). This allows us to further simplify the equations of motion by plugging the equations of the input or output modes (\ref{eq:input-output-modes}) into (\ref{eq:resonator-modes}) to yield
\begin{subequations}
\begin{align}
\dot{\hat{b}}_j(t) &= -i\tilde{\omega}_j \hat{b}_j(t) - i\sqrt{\beta_j}\hat{p} _{\text{in}}(t) - \sum_{n \in \{r,1,2\}} \sqrt{\beta_j \beta_n} \hat{b}_n(t) - i\hat{\phi}_j(t) \\
\dot{\hat{b}}_j(t) &= -i\tilde{\omega}_j \hat{b}_j(t) - i\sqrt{\beta_j}\hat{p}_{\text{out}}(t) + \sum_{n \in \{r,1,2\}} \sqrt{\beta_j \beta_n} \hat{b}_n(t) - i\hat{\phi}_j(t).
\end{align}
\end{subequations}
where \(\hat{\phi}_j(t)\) is defined as
\begin{equation}
\hat{\phi}_j(t) = \begin{cases} g_1^* \hat{b}_1(t) + g_2^* \hat{b}_2(t) & ; j=r \\ g_j \hat{b}_r(t) & ; j=1,2 \end{cases}.
\end{equation}
Taking the Fourier transform of these expanded equations gives
\begin{subequations}
\begin{align}
i(\tilde{\omega}j \pm \omega)\hat{b}_j(\omega) + i\sqrt{\beta_j}\hat{p}_{\text{in}}(\omega) + \sum_{n}& \sqrt{\beta_j \beta_n} \hat{b}_n(\omega) + i\hat{\phi}_j(\omega) = 0 \\
i(\tilde{\omega}j \pm \omega)\hat{b}_j(\omega) + i\sqrt{\beta_j}\hat{p}_{\text{out}}(\omega) - \sum_{n}& \sqrt{\beta_j \beta_n} \hat{b}_n(\omega) + i\hat{\phi}_j(\omega) = 0.
\end{align}
\end{subequations}
Finally, by defining the \(S_{21}\) parameter as \(S_{21}=\hat{p}_{\text{out}}/\hat{p}_{\text{in}}-1\) and substituting the derived expressions, we arrive at the final result:
\begin{equation}
S_{21}(\omega) = \frac{2}{i} \sum_n \sqrt{\beta_n}\frac{ \hat{b}_n(\omega)}{\hat{p}_{\text{in}}(\omega)},
\end{equation}
which matches Equation \ref{eq:s21-analytical} precisely.

\medskip

%

\bibliography{bibliography}

\begin{thebibliography}{36}
\providecommand{\natexlab}[1]{#1}
\providecommand{\url}[1]{\texttt{#1}}
\expandafter\ifx\csname urlstyle\endcsname\relax
  \providecommand{\doi}[1]{doi: #1}\else
  \providecommand{\doi}{doi: \begingroup \urlstyle{rm}\Url}\fi

\bibitem[Wagle et~al.(2024)Wagle, Rai, Kaffash, and Jungfleisch]{Wagle_2024}
Dinesh Wagle, Anish Rai, Mojtaba~T Kaffash, and M~Benjamin Jungfleisch.
\newblock Controlling magnon-photon coupling in a planar geometry.
\newblock \emph{Journal of Physics: Materials}, 7\penalty0 (2):\penalty0 025005, February 2024.
\newblock ISSN 2515-7639.
\newblock \doi{10.1088/2515-7639/ad2984}.
\newblock URL \url{http://dx.doi.org/10.1088/2515-7639/ad2984}.

\bibitem[Zhao et~al.(2022)Zhao, Peng, Yang, Chao, Li, Wang, and Zhou]{zhao2022}
Chengsong Zhao, Rui Peng, Zhen Yang, Shilei Chao, Chong Li, Zhihai Wang, and Ling Zhou.
\newblock Nonreciprocal amplification in a cavity magnonics system.
\newblock \emph{Phys. Rev. A}, 105:\penalty0 023709, Feb 2022.
\newblock \doi{10.1103/PhysRevA.105.023709}.
\newblock URL \url{https://link.aps.org/doi/10.1103/PhysRevA.105.023709}.

\bibitem[Zhu et~al.(2020)Zhu, Han, Zou, Xu, and Tang]{zhu2020}
Na~Zhu, Xu~Han, Chang-Ling Zou, Mingrui Xu, and Hong~X. Tang.
\newblock Magnon-photon strong coupling for tunable microwave circulators.
\newblock \emph{Phys. Rev. A}, 101:\penalty0 043842, Apr 2020.
\newblock \doi{10.1103/PhysRevA.101.043842}.
\newblock URL \url{https://link.aps.org/doi/10.1103/PhysRevA.101.043842}.

\bibitem[Zhang et~al.(2021)Zhang, Jia, Shi, Jiang, Xue, Ong, and Chai]{zhang2021}
Chi Zhang, Chenglong Jia, Yongzhang Shi, Changjun Jiang, Desheng Xue, C.~K. Ong, and Guozhi Chai.
\newblock Nonreciprocal multimode and indirect couplings in cavity magnonics.
\newblock \emph{Phys. Rev. B}, 103:\penalty0 184427, May 2021.
\newblock \doi{10.1103/PhysRevB.103.184427}.
\newblock URL \url{https://link.aps.org/doi/10.1103/PhysRevB.103.184427}.

\bibitem[Fan et~al.(2025)Fan, Zuo, Li, and Li]{fan2025}
Zhiyuan Fan, Xuan Zuo, Haotian Li, and Jie Li.
\newblock Nonreciprocal entanglement in cavity magnomechanics exploiting chiral cavity-magnon coupling.
\newblock \emph{Fundamental Research}, 2025.
\newblock ISSN 2667-3258.
\newblock \doi{https://doi.org/10.1016/j.fmre.2025.02.012}.
\newblock URL \url{https://www.sciencedirect.com/science/article/pii/S2667325825000950}.

\bibitem[Li et~al.(2023)Li, Draher, Comstock, Xiong, Haque, Easy, Qian, Polakovic, Pearson, Divan, Zuo, Zhang, Welp, Kwok, Hoffmann, Luther, Beard, Sun, Zhang, and Novosad]{li2023tunablemagnonphotoncouplingmagnon}
Yi~Li, Timothy Draher, Andrew~H. Comstock, Yuzan Xiong, Md~Azimul Haque, Elham Easy, Jiangchao Qian, Tomas Polakovic, John~E. Pearson, Ralu Divan, Jian-Min Zuo, Xian Zhang, Ulrich Welp, Wai-Kwong Kwok, Axel Hoffmann, Joseph~M. Luther, Matthew~C. Beard, Dali Sun, Wei Zhang, and Valentine Novosad.
\newblock Probing intrinsic magnon bandgap in a layered hybrid perovskite antiferromagnet by a superconducting resonator.
\newblock \emph{Phys. Rev. Res.}, 5:\penalty0 043031, Oct 2023.
\newblock \doi{10.1103/PhysRevResearch.5.043031}.
\newblock URL \url{https://link.aps.org/doi/10.1103/PhysRevResearch.5.043031}.

\bibitem[Ghirri et~al.(2023)Ghirri, Bonizzoni, Maksutoglu, Mercurio, Di~Stefano, Savasta, and Affronte]{Ghirri2023}
Alberto Ghirri, Claudio Bonizzoni, Maksut Maksutoglu, Alberto Mercurio, Omar Di~Stefano, Salvatore Savasta, and Marco Affronte.
\newblock Ultrastrong magnon-photon coupling achieved by magnetic films in contact with superconducting resonators.
\newblock \emph{Phys. Rev. Appl.}, 20:\penalty0 024039, Aug 2023.
\newblock \doi{10.1103/PhysRevApplied.20.024039}.
\newblock URL \url{https://link.aps.org/doi/10.1103/PhysRevApplied.20.024039}.

\bibitem[Kim et~al.(2024)Kim, Kim, Kim, Jeon, and Kim]{Kim2024}
Junyoung Kim, Bosung Kim, Bojong Kim, Haechan Jeon, and Sang-Koog Kim.
\newblock Magnetic-field controlled on-off switchable non-reciprocal negative refractive index in non-hermitian photon-magnon hybrid systems.
\newblock \emph{Nature Communications}, 15\penalty0 (1):\penalty0 9014, Oct 2024.
\newblock ISSN 2041-1723.
\newblock \doi{10.1038/s41467-024-53328-9}.
\newblock URL \url{https://doi.org/10.1038/s41467-024-53328-9}.

\bibitem[Freeman et~al.(2025)Freeman, Youel, Budniak, Xue, De~Libero, Thomson, Bosman, Eda, Kurebayashi, and Cubukcu]{freeman2025}
Charlie W.~F. Freeman, Harry Youel, Adam~K. Budniak, Zekun Xue, Henry De~Libero, Thomas Thomson, Michel Bosman, Goki Eda, Hidekazu Kurebayashi, and Murat Cubukcu.
\newblock Tunable ultrastrong magnon–magnon coupling approaching the deep-strong regime in a van der waals antiferromagnet.
\newblock \emph{ACS Nano}, 19\penalty0 (16):\penalty0 16024--16031, 2025.
\newblock \doi{10.1021/acsnano.5c02576}.
\newblock URL \url{https://doi.org/10.1021/acsnano.5c02576}.
\newblock PMID: 40244900.

\bibitem[Cosset-Ch\'eneau et~al.(2024)Cosset-Ch\'eneau, Tirion, Wei, Ben~Youssef, and van Wees]{cosset2024}
M.~Cosset-Ch\'eneau, S.~H. Tirion, X.-Y. Wei, J.~Ben~Youssef, and B.~J. van Wees.
\newblock Nonreciprocal transport of thermally generated magnons.
\newblock \emph{Phys. Rev. B}, 110:\penalty0 214418, Dec 2024.
\newblock \doi{10.1103/PhysRevB.110.214418}.
\newblock URL \url{https://link.aps.org/doi/10.1103/PhysRevB.110.214418}.

\bibitem[Chakraborty and Das(2023)]{chakraborty2023}
Subhadeep Chakraborty and Camelia Das.
\newblock Nonreciprocal magnon-photon-phonon entanglement in cavity magnomechanics.
\newblock \emph{Phys. Rev. A}, 108:\penalty0 063704, Dec 2023.
\newblock \doi{10.1103/PhysRevA.108.063704}.
\newblock URL \url{https://link.aps.org/doi/10.1103/PhysRevA.108.063704}.

\bibitem[Nongthombam et~al.(2023)Nongthombam, Gupta, and Sarma]{nongthombam2023}
Roson Nongthombam, Pooja~Kumari Gupta, and Amarendra~K. Sarma.
\newblock Quantum transduction of a superconducting qubit in an electro-optomechanical and an electro-optomagnonical system.
\newblock \emph{Phys. Rev. A}, 108:\penalty0 043501, Oct 2023.
\newblock \doi{10.1103/PhysRevA.108.043501}.
\newblock URL \url{https://link.aps.org/doi/10.1103/PhysRevA.108.043501}.

\bibitem[Hou and Liu(2019)]{hou2019}
Justin~T. Hou and Luqiao Liu.
\newblock Strong coupling between microwave photons and nanomagnet magnons.
\newblock \emph{Phys. Rev. Lett.}, 123:\penalty0 107702, Sep 2019.
\newblock \doi{10.1103/PhysRevLett.123.107702}.
\newblock URL \url{https://link.aps.org/doi/10.1103/PhysRevLett.123.107702}.

\bibitem[Li et~al.(2019)Li, Polakovic, Wang, Xu, Lendinez, Zhang, Ding, Khaire, Saglam, Divan, Pearson, Kwok, Xiao, Novosad, Hoffmann, and Zhang]{li2019}
Yi~Li, Tomas Polakovic, Yong-Lei Wang, Jing Xu, Sergi Lendinez, Zhizhi Zhang, Junjia Ding, Trupti Khaire, Hilal Saglam, Ralu Divan, John Pearson, Wai-Kwong Kwok, Zhili Xiao, Valentine Novosad, Axel Hoffmann, and Wei Zhang.
\newblock Strong coupling between magnons and microwave photons in on-chip ferromagnet-superconductor thin-film devices.
\newblock \emph{Phys. Rev. Lett.}, 123:\penalty0 107701, Sep 2019.
\newblock \doi{10.1103/PhysRevLett.123.107701}.
\newblock URL \url{https://link.aps.org/doi/10.1103/PhysRevLett.123.107701}.

\bibitem[Xu et~al.(2024)Xu, Cheung, Cormode, Puel, Pal, Yusuf, Chilcote, Flatté, Johnston-Halperin, and Fuchs]{Xu2024}
Qin Xu, Hil Fung~Harry Cheung, Donley~S. Cormode, Tharnier~O. Puel, Srishti Pal, Huma Yusuf, Michael Chilcote, Michael~E. Flatté, Ezekiel Johnston-Halperin, and Gregory~D. Fuchs.
\newblock Strong photon-magnon coupling using a lithographically defined organic ferrimagnet.
\newblock \emph{Advanced Science}, 11\penalty0 (14):\penalty0 2310032, 2024.
\newblock \doi{https://doi.org/10.1002/advs.202310032}.
\newblock URL \url{https://advanced.onlinelibrary.wiley.com/doi/abs/10.1002/advs.202310032}.

\bibitem[Jeon et~al.(2024)Jeon, Kim, Kim, Bhoi, and Kim]{Jeon2024}
Haechan Jeon, Bojong Kim, Junyoung Kim, Biswanath Bhoi, and Sang-Koog Kim.
\newblock Anomalous coherent and dissipative coupling in dual photon-magnon hybrid resonators.
\newblock \emph{Scientific Reports}, 14\penalty0 (1):\penalty0 13581, Jun 2024.
\newblock ISSN 2045-2322.
\newblock \doi{10.1038/s41598-024-64315-x}.
\newblock URL \url{https://doi.org/10.1038/s41598-024-64315-x}.

\bibitem[Bhoi et~al.(2017)Bhoi, Kim, Kim, Cho, and Kim]{Bhoi2017}
Biswanath Bhoi, Bosung Kim, Junhoe Kim, Young-Jun Cho, and Sang-Koog Kim.
\newblock Robust magnon-photon coupling in a planar-geometry hybrid of inverted split-ring resonator and yig film.
\newblock \emph{Scientific Reports}, 7\penalty0 (1):\penalty0 11930, Sep 2017.
\newblock ISSN 2045-2322.
\newblock \doi{10.1038/s41598-017-12215-8}.
\newblock URL \url{https://doi.org/10.1038/s41598-017-12215-8}.

\bibitem[Andrich et~al.(2017)Andrich, de~las Casas, Liu, Bretscher, Berman, Heremans, Nealey, and Awschalom]{Andrich_2017}
Paolo Andrich, Charles~F. de~las Casas, Xiaoying Liu, Hope~L. Bretscher, Jonson~R. Berman, F.~Joseph Heremans, Paul~F. Nealey, and David~D. Awschalom.
\newblock Long-range spin wave mediated control of defect qubits in nanodiamonds.
\newblock \emph{npj Quantum Information}, 3\penalty0 (1), July 2017.
\newblock ISSN 2056-6387.
\newblock \doi{10.1038/s41534-017-0029-z}.
\newblock URL \url{http://dx.doi.org/10.1038/s41534-017-0029-z}.

\bibitem[Hyde et~al.(2016)Hyde, Bai, Harder, Match, and Hu]{cannot-g3}
Paul Hyde, Lihui Bai, Michael Harder, Christophe Match, and Can-Ming Hu.
\newblock Indirect coupling between two cavity modes via ferromagnetic resonance.
\newblock \emph{Applied Physics Letters}, 109\penalty0 (15):\penalty0 152405, 10 2016.
\newblock ISSN 0003-6951.
\newblock \doi{10.1063/1.4964602}.
\newblock URL \url{https://doi.org/10.1063/1.4964602}.

\bibitem[Grigoryan and Xia(2019)]{grigoryan2019}
Vahram~L. Grigoryan and Ke~Xia.
\newblock Cavity-mediated dissipative spin-spin coupling.
\newblock \emph{Phys. Rev. B}, 100:\penalty0 014415, Jul 2019.
\newblock \doi{10.1103/PhysRevB.100.014415}.
\newblock URL \url{https://link.aps.org/doi/10.1103/PhysRevB.100.014415}.

\bibitem[Morris et~al.(2017)Morris, van Loo, Kosen, and Karenowska]{morris2017}
R.~G.~E. Morris, A.~F. van Loo, S.~Kosen, and A.~D. Karenowska.
\newblock Strong coupling of magnons in a yig sphere to photons in a planar superconducting resonator in the quantum limit.
\newblock \emph{Scientific Reports}, 7\penalty0 (1):\penalty0 11511, Sep 2017.
\newblock ISSN 2045-2322.
\newblock \doi{10.1038/s41598-017-11835-4}.
\newblock URL \url{https://doi.org/10.1038/s41598-017-11835-4}.

\bibitem[Zhang et~al.(2016)Zhang, Zou, Jiang, and Tang]{zhang2016}
Xufeng Zhang, Chang-Ling Zou, Liang Jiang, and Hong~X. Tang.
\newblock Cavity magnomechanics.
\newblock \emph{Science Advances}, 2\penalty0 (3):\penalty0 e1501286, 2016.
\newblock \doi{10.1126/sciadv.1501286}.
\newblock URL \url{https://www.science.org/doi/abs/10.1126/sciadv.1501286}.

\bibitem[Bourhill et~al.(2016)Bourhill, Kostylev, Goryachev, Creedon, and Tobar]{bourhill2016}
J.~Bourhill, N.~Kostylev, M.~Goryachev, D.~L. Creedon, and M.~E. Tobar.
\newblock Ultrahigh cooperativity interactions between magnons and resonant photons in a yig sphere.
\newblock \emph{Phys. Rev. B}, 93:\penalty0 144420, Apr 2016.
\newblock \doi{10.1103/PhysRevB.93.144420}.
\newblock URL \url{https://link.aps.org/doi/10.1103/PhysRevB.93.144420}.

\bibitem[Dilawaiz et~al.(2024)Dilawaiz, Qamar, and Irfan]{Dilawaiz_2024}
Dilawaiz, Shahid Qamar, and Muhammad Irfan.
\newblock Entangled atomic ensemble and an yttrium-iron-garnet sphere in coupled microwave cavities.
\newblock \emph{Physical Review A}, 109\penalty0 (4), April 2024.
\newblock ISSN 2469-9934.
\newblock \doi{10.1103/physreva.109.043708}.
\newblock URL \url{http://dx.doi.org/10.1103/PhysRevA.109.043708}.

\bibitem[Inman et~al.(2022)Inman, Xiong, Bidthanapally, Louis, Tyberkevych, Qu, Sklenar, Novosad, Li, Zhang, and Zhang]{inman2022}
Jerad Inman, Yuzan Xiong, Rao Bidthanapally, Steven Louis, Vasyl Tyberkevych, Hongwei Qu, Joseph Sklenar, Valentine Novosad, Yi~Li, Xufeng Zhang, and Wei Zhang.
\newblock Hybrid magnonics for short-wavelength spin waves facilitated by a magnetic heterostructure.
\newblock \emph{Phys. Rev. Appl.}, 17:\penalty0 044034, Apr 2022.
\newblock \doi{10.1103/PhysRevApplied.17.044034}.
\newblock URL \url{https://link.aps.org/doi/10.1103/PhysRevApplied.17.044034}.

\bibitem[Santos and van Wees(2023)]{Santos2023}
Obed~Alves Santos and Bart~J. van Wees.
\newblock Magnon confinement in an all-on-chip yig cavity resonator using hybrid yig/py magnon barriers.
\newblock \emph{Nano Letters}, 23\penalty0 (20):\penalty0 9303--9309, 2023.
\newblock \doi{10.1021/acs.nanolett.3c02388}.
\newblock URL \url{https://doi.org/10.1021/acs.nanolett.3c02388}.
\newblock PMID: 37819876.

\bibitem[Xiong et~al.(2020)Xiong, Li, Hammami, Bidthanapally, Sklenar, Zhang, Qu, Srinivasan, Pearson, Hoffmann, Novosad, and Zhang]{Xiong2020}
Yuzan Xiong, Yi~Li, Mouhamad Hammami, Rao Bidthanapally, Joseph Sklenar, Xufeng Zhang, Hongwei Qu, Gopalan Srinivasan, John Pearson, Axel Hoffmann, Valentine Novosad, and Wei Zhang.
\newblock Probing magnon--magnon coupling in exchange coupled {Y${}_3$Fe${}_5$O${}_{12}$/Permalloy} bilayers with magneto-optical effects.
\newblock \emph{Scientific Reports}, 10\penalty0 (1):\penalty0 12548, Jul 2020.
\newblock ISSN 2045-2322.
\newblock \doi{10.1038/s41598-020-69364-6}.
\newblock URL \url{https://doi.org/10.1038/s41598-020-69364-6}.

\bibitem[Walls and Milburn(2007)]{walls-2007}
D.F. Walls and Gerard~J. Milburn.
\newblock \emph{{Quantum Optics}}.
\newblock Springer Science \& Business Media, 12 2007.

\bibitem[Shrivastava et~al.(2024{\natexlab{a}})Shrivastava, Sahu, Bhoi, and Singh]{shrivastava-2024}
Kuldeep~Kumar Shrivastava, Ansuman Sahu, Biswanath Bhoi, and Rajeev Singh.
\newblock {Unveiling photon-photon coupling induced transparency and absorption}.
\newblock \emph{Journal of Physics D Applied Physics}, 57\penalty0 (46):\penalty0 465305, 8 2024{\natexlab{a}}.
\newblock \doi{10.1088/1361-6463/ad6613}.
\newblock URL \url{https://doi.org/10.1088/1361-6463/ad6613}.

\bibitem[Grigoryan and Xiao(2013)]{Grigoryan_2013}
Vahram Grigoryan and Jiang Xiao.
\newblock Dynamical spin-spin coupling of quantum dots.
\newblock \emph{Europhysics Letters}, 104\penalty0 (1):\penalty0 17008, oct 2013.
\newblock \doi{10.1209/0295-5075/104/17008}.
\newblock URL \url{https://dx.doi.org/10.1209/0295-5075/104/17008}.

\bibitem[Schrieffer and Wolff(1966)]{schrieffer1966}
J.~R. Schrieffer and P.~A. Wolff.
\newblock Relation between the anderson and kondo hamiltonians.
\newblock \emph{Phys. Rev.}, 149:\penalty0 491--492, Sep 1966.
\newblock \doi{10.1103/PhysRev.149.491}.
\newblock URL \url{https://link.aps.org/doi/10.1103/PhysRev.149.491}.

\bibitem[Kittel(1948)]{kittel1948}
Charles Kittel.
\newblock On the theory of ferromagnetic resonance absorption.
\newblock \emph{Phys. Rev.}, 73:\penalty0 155--161, Jan 1948.
\newblock \doi{10.1103/PhysRev.73.155}.
\newblock URL \url{https://link.aps.org/doi/10.1103/PhysRev.73.155}.

\bibitem[Bernier et~al.(2018)Bernier, T\'oth, Feofanov, and Kippenberg]{bernier2018}
N.~R. Bernier, L.~D. T\'oth, A.~K. Feofanov, and T.~J. Kippenberg.
\newblock Level attraction in a microwave optomechanical circuit.
\newblock \emph{Phys. Rev. A}, 98:\penalty0 023841, Aug 2018.
\newblock \doi{10.1103/PhysRevA.98.023841}.
\newblock URL \url{https://link.aps.org/doi/10.1103/PhysRevA.98.023841}.

\bibitem[Metelmann and Clerk(2014)]{metelmann2014}
A.~Metelmann and A.~A. Clerk.
\newblock Quantum-limited amplification via reservoir engineering.
\newblock \emph{Phys. Rev. Lett.}, 112:\penalty0 133904, Apr 2014.
\newblock \doi{10.1103/PhysRevLett.112.133904}.
\newblock URL \url{https://link.aps.org/doi/10.1103/PhysRevLett.112.133904}.

\bibitem[Rao et~al.(2020)Rao, Wang, Yang, Yu, Gui, Fan, Xue, and Hu]{rao2020}
J.~W. Rao, Y.~P. Wang, Y.~Yang, T.~Yu, Y.~S. Gui, X.~L. Fan, D.~S. Xue, and C.-M. Hu.
\newblock Interactions between a magnon mode and a cavity photon mode mediated by traveling photons.
\newblock \emph{Phys. Rev. B}, 101:\penalty0 064404, Feb 2020.
\newblock \doi{10.1103/PhysRevB.101.064404}.
\newblock URL \url{https://link.aps.org/doi/10.1103/PhysRevB.101.064404}.

\bibitem[Shrivastava et~al.(2024{\natexlab{b}})Shrivastava, Ketkar, Bhoi, and Singh]{shrivastava2024emergencecouplinginducedtransparency}
Kuldeep~Kumar Shrivastava, Moulik~Deviprasad Ketkar, Biswanath Bhoi, and Rajeev Singh.
\newblock Emergence of coupling induced transparency by tuning purely dissipative couplings, 2024{\natexlab{b}}.
\newblock URL \url{https://arxiv.org/abs/2409.12577}.

\end{thebibliography}


\end{document}